\documentclass[letterpaper, 10 pt, conference]{ieeeconf}  

\IEEEoverridecommandlockouts                              
\usepackage{graphicx}

\title{\LARGE \bf
Exposing Vulnerabilities in Mobile Networks: A Mobile Data Consumption Attack}

\author{Dean Wasil$^{1}$, Omar Nakhila$^{2}$, Salih Safa Bacanli$^{3}$, Cliff Zou$^{3}$ and Damla Turgut$^{3}$ \\
$^{1}$ Department of Computer Science, Mount Vernon Nazarene University \\
$^{2}$ Department of Electrical and Computer Engineering, University of Central Florida \\
$^{3}$ Department of Computer Science, University of Central Florida 
}

\begin{document}

\maketitle
\thispagestyle{empty}
\pagestyle{empty}

\begin{abstract}
Smartphone carrier companies rely on mobile networks for keeping an accurate record of customer data usage for billing purposes. In this paper, we present a vulnerability that allows an attacker to force the victim's smartphone to consume data through the cellular network by starting the data download on the victim's cell phone without the victim's knowledge. The attack is based on switching the victim's smartphones from the Wi-Fi network to the cellular network while downloading a large data file. This attack has been implemented in real-life scenarios where the test's outcomes demonstrate that the attack is feasible and that mobile networks do not record customer data usage accurately.
\end{abstract}

\begin{keywords}
WiFi Security; Evil Twin Attack; Man in the Middle Attack; Deauthentication Attack; Captive Portal; Mobile Data Consumption.
\end{keywords}

\section{INTRODUCTION}
Cell phones are becoming a necessary piece of equipment in our lives. In 2016, there were seven billion active cellular telephone subscriptions worldwide with three and half billion cell phones that have a subscription to access the Internet \cite{c5}. To reduce congestion on the cellular network, most cell phone ISP carriers throttle customer data after exceeding a certain data limit or impose a monthly limit data cap. 
\par
On the other hand, as a complementary service, coffee shops, fast food restaurants, and airports provide free Wi-Fi network access to their customers. These open access networks provide budget-friendly Internet access which helps offload data traffic from the cellular network \cite{c6}\cite{c7}. However, for their ease of access, these types of networks are insecure in terms of lacking authentication and encryption. Instead, when the customer first accesses the Wi-Fi network, he or she must agree to “the Public Wi-Fi Terms and Conditions,” where the ISP provider claims no responsibility for the customer information security/privacy \cite{c8}. 
\par
\begin{figure}[!t]
      \centering
      \includegraphics[width=\linewidth]{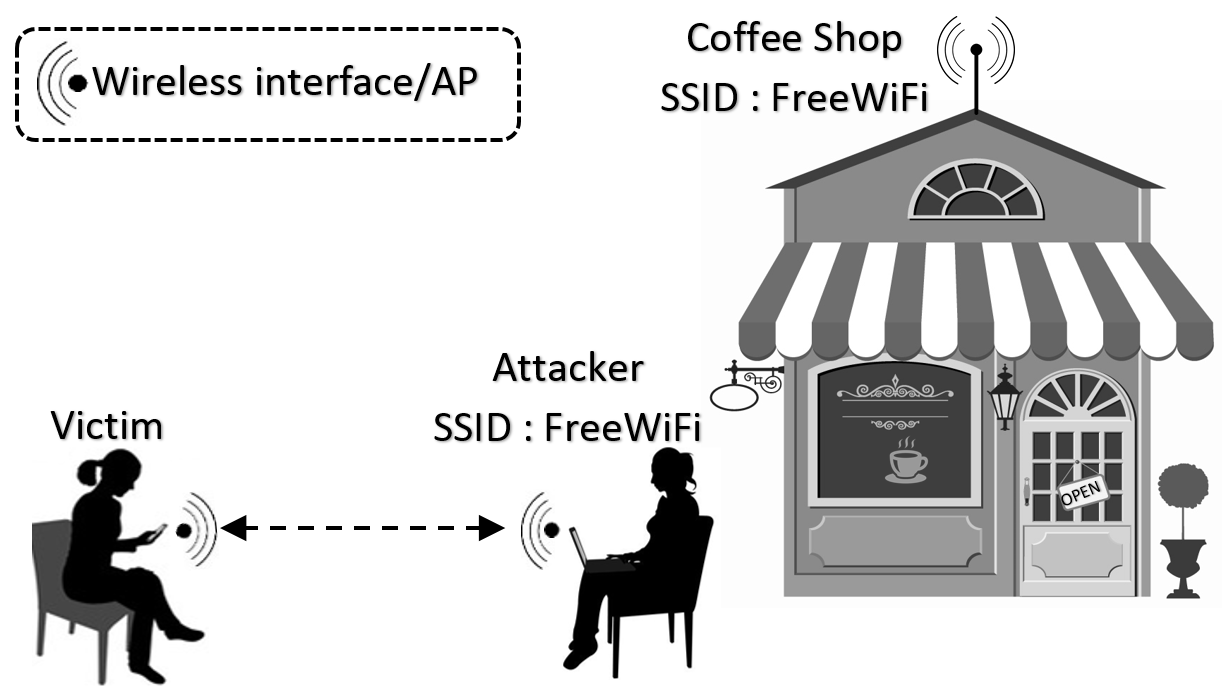}
      \caption{Illustration of an Evil Twin Attack . The attacker can successfully lure a victim into connecting to a fake access point instead of the legitimate access point when it provides a stronger/better signal to those customers.}
      \label{figure0}
   \end{figure}

The absence of wireless security in open access Wi-Fi networks provides tempting environments for attackers \cite{c3}\cite{c4}\cite{c9}\cite{c10}. An Evil Twin Attack (ETA) can be initiated by an attacker to impersonate the role of a legitimate Wi-Fi access point (illustrated in Fig. \ref{figure0}). Such an impersonation is simple since an open Wi-Fi network can only be recognized by its MAC address and Service Set Identifier (SSID). Furthermore, the attacker's fake access point (AP) may provide a better and more powerful signal to the victim in which case it will cause the victim to connect to the attacker's network instead of the legitimate network. Such a switch is automatic and can happen without the victim's intervention \cite{c18}. 
\par 
After the victim connects to the attacker's AP, the attacker can snoop the victim's wireless data and apply a Man In The Middle Attacks (MIMA). For example, DNS spoofing attack \cite{c19} is a popular MIMA that allows the attacker to response to DNS query coming from the victim. The attack can redirect the victim web browser to a malicious website rather than the legitimate one by sending her a wrong IP address.
\par
In this paper, we improved wireless network security by:
\begin{itemize}
\item Introducing a new vulnerability that depletes customer mobile data quota. The attack targets customers that use free open Wi-Fi networks instead of their cellular data connection.
\item Our proposed attack was implemented and evaluated in a real-life environment. 
\end{itemize}

The paper is organized as follows. Section \ref{relatedwork} discusses related works. The design of the new mobile data consumption attack is presented in Section \ref{design}. Then, our attack results are shown in Section \ref{results}. Finally, we present our conclusions in section \ref{con}. Throughout the paper, we refer to the attacker as an individual or group of individuals.

\section{RELATED WORK} \label{relatedwork}
A mobile data consumption attack can drain a victim's data cap in a short period of time. Such an attack can prevent the victim from accessing the Internet after reaching his or her monthly allocated data limit, which leads to denial of service. If the victim does not have a data cap, he or she will be continuously charged for the data used by the attack. Also, keeping the mobile device transmitting/receiving data results in battery power consumption.
\par   
Stealth spam attacks can abuse the fact that many connections that are formed do not tell the network that they are closing \cite{c1}. Thus an attacker can use a connection made by the victim that the network still thinks is open, even though the phone had closed it. The attacker can then send data over this connection as a spam attack, which consumes the victim's data.
\par
Other similar attacks are introduced in \cite{c2}. The first is the “cloak-and-dagger spamming attack” where the victim's data is consumed by either spoofing the victim's IP address and using data as the victim, or by sending an MMS message which opens up a connection to spam the victim over, using up the victim's data. The second attack in \cite{c2} is the “hit-but-no-touch attack” where data packets are sent to the victim with a shortened time-to-live value. The packets then pass through a mobile network's billing system, but never make it to the victim, thus invisibly using the victim's data.
\par
Mobile billing vulnerability based on TCP packet re-transmission is presented in \cite{c10}. Many mobile service providers, such as AT\&T, Verizon, T-Mobile, and Sprint, bill customers based on the total amount of data traffic that has been sent and received to the Internet, including retransmitted data packets. An attacker can force the victim to consume more data by increasing the number of TCP data packets retransmitted.
\par
The attacker sends a text message to the victim with a link to a malicious server. When the victim opens the link, a TCP connection between the victim's cell phone and the remote server will be established. The connection is based on TCP protocol, and thus it can keep forcing the victim to retransmit TCP packets. TCP packet retransmission can be initiated when three acknowledgment packets were received to the same TCP sequence number or by the timeout of the TCP connection.
\par
A similar attack on a mobile billing system is when the attacker sends a spoofed IP packet to an external server \cite{c11}. The remote server uses the victim's spoofed IP address to start a network connection to the victim's cell phone. The remote server then sends a large amount of data to the mobile network that will be directed to the victim cell phone. 
\par
\begin{figure}[!t]
      \centering
      \includegraphics[scale=0.34]{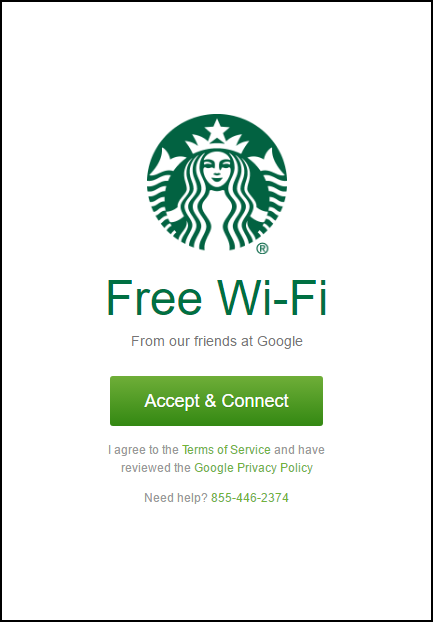}
      \caption{A common Starbucks captive portal with an injected malicious script. Once this page opens, a download begins in the victim's smartphone's background.}
      \label{figure1}
\end{figure}
All the previous attacks targeted victims that use a private IP address. The impact of these attacks can be intensified when the victim is using a public IP address \cite{c12}. With the spread of IPv6, cell phone devices will have direct access to the Internet. In this situation, the attacker can directly send data packets to the victim that deplete cell phone data quota.
\par
In this paper, we present a new attack that exploits the inaccuracy of mobile network billing systems.

\section{Proposed Mobile Data Consumption Attack}\label{design}
\subsection{Preliminaries}
Here are the terms we use throughout the paper:

\begin{itemize}
\item {\bf Attacker:} the individual or group of individuals carrying out the attack and the attacker's equipment used, such as a laptop and software.
\item {\bf Victim:} the individual that is being attacked and the device that is being attacked, such as the victim's smartphone.
\item {\bf Captive Portal:} a web page that network users are redirected to in order to accept these network usage conditions or similar terms. They are often used in coffee shops, fast food restaurants, and airports. They can be seen directly after a user connects to the network, as the user will be redirected to the captive portal upon attempting to use the Internet. If the customer doesn't accept the terms and conditions on the captive portal, he or she will be denied Internet access to the free Wi-Fi network.
\end{itemize}

Our mobile data consumption attack is designed for use where there is a nearby public Wi-Fi hotspot, such as a coffee shop, hotel, fast food restaurant, or store that has a captive portal.   
The attacker can target a victim or set of victims at the Wi-Fi network. If the attacker targeted a particular victim, the attacker could wait until the victim enters an area with a nearby public Wi-Fi hotspot. 

In the current version of the attack, any customer connected to the open Wi-Fi network is a potential target for our proposed attack. However, not every person will be attacked. This work focuses on attacking one victim, as attacking a set of victims requires running the attack multiple times, once on each victim. Selecting the victim or setting up the attack can happen in either order. If the attacker chooses to attack an individual that happens to be at the location, it is recommended that the attacker sets up the attack before choosing a random victim. If the attacker has prior knowledge of a particular victim's plan to go to a certain location, the attacker can setup the attack before the victim arrives. Having the attack setup before the victim arrives or before the victim is chosen increases the chance of the attack's success. 


\subsection{Design}
The proposed attack in this paper is designed based on the following three attacks:
\begin{itemize}
\item The attacker creates a fake web server that serves a captive portal web page that is similar to the original Wi-Fi network. The captive portal web page includes a malicious code that forces the victim to download a large data file from the Internet.
\item Using an Evil Twin Attack, the attacker lures the victim to switch to the fake network. Such a switch can also happen in an automatic manner especially when the attacker AP is near to the victim's location. After the victim connects to the fake network, the attacker can spoof the victim's DNS request by sending the malicious captive portal web page whenever the victim requests an URL.
\item To make sure the victim is only downloading data through the cellular network, deauthentication attacks target the victim's smartphone preventing him or her from connection to any Wi-Fi network after the captive portal is delivered. Deauthentication is easy to implement because 802.11 WLAN management wireless frames are sent without any protection \cite{c13}.

\end{itemize}
      \begin{figure}[!t]
      \centering
      \includegraphics[scale=0.46]{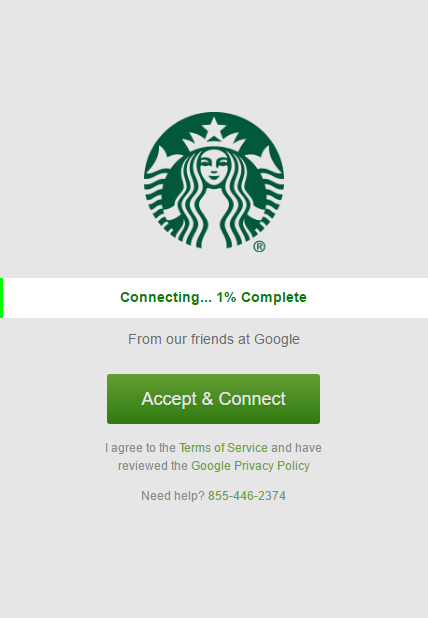}
      \caption{The overlay displayed by the malicious script. The loading bar moves to 100\% using a logarithmic progression designed to keep the victim on the page so that the attacker can perform the attack.}
      \label{figure2}
   \end{figure}
   
\subsection{Implementation}
A laptop with an off-shelf network interface card is used in our attack proposal. Linux operating system was used to implement all the attacks illustrated in the design section. First, the attacker starts an Evil Twin Attack on the victim. We can predict which open Wi-Fi network the victim is connected to, based on his or her current location. For instance, if the victim is in a coffee shop, it is likely that the victim is connected to the coffee shop's public Wi-Fi. The attacker must connect to the open Wi-Fi network beforehand and capture the captive portal used on the network. A captive portal can be captured by connecting to the original Wi-Fi network and accessing the Internet using a free web browser, such as Chrome. The original Wi-Fi network sends the captive portal web page to the attacker which is downloaded and used to create the malicious captive portal.
      \begin{figure}[!t]
      \centering
      \includegraphics[scale=0.46]{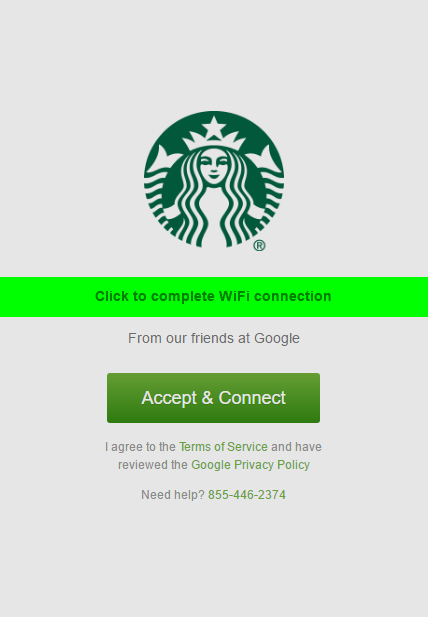}
      \caption{The captive portal after the loading bar finishes. When the victim clicks on the page, the victim is redirected to her desired URL, but this page is left open in another tab, downloading in the background.}
      \label{figure3}
   \end{figure}

\par
Not every captive portal that is downloaded will be displayed to the victim exactly the way it appears on the Wi-Fi. Because of this, the attacker may have to manually adjust the captive portal's code in order to make the downloaded captive portal look very similar to the original captive portal. The closer the downloaded captive portal looks to the original, the better the chance is that a victim will not notice that the downloaded captive portal is not the original one.
\par
Once a downloaded copy of the captive portal is obtained, the attacker needs to inject malicious code into the captive portal. The code has been written for this attack, and it can be injected into most captive portals without altering the integrity of the web page. The code does not change how a captive portal looks. However, the new malicious captive portal adds functionality that causes the web page to download data in the background. Functionality is also added that causes the web page to display an overlay upon clicking the button on the captive portal that allows a user to connect to the network. 
\par
The overlay is designed to give time for the attacker to conduct the attack while attempting to keep the victim on the captive portal and not realize that the attack is occurring. Also, the captive portal is designed to cause the victim to leave the web page open for an extended period of time by redirecting the victim away from the captive portal, leaving it open and downloading in another tab. The code has also been designed to bypass most pop-up blockers which are built into smartphone browsers. An example of a malicious captive portal is shown in Figure~\ref{figure1} along with examples of the overlay shown in Figure~\ref{figure2} and Figure~\ref{figure3}.

\section{Results} \label{results}
After the malicious captive portal is implemented, the attacker is ready to start our proposed attack. The attacker will begin the mobile data consumption attack by initiating an Evil Twin Attack \cite{c3}\cite{c4} in which the attacker's laptop Wi-Fi impersonates the original access point that the victim is connected to. The program Airedump-ng can be used to find information about the AP the victim is connected to, such as the AP's MAC address and the AP's channel. The program Airemon-ng \cite{c14} can be used to mount the attacker's laptop wireless card into monitor mode to prepare the card to be used as an AP. The program Airebase-ng \cite{c14} can then be used to make the card work as an AP with the same MAC address and name of the public AP. The attacker should now have a replica of the public AP running on his or her laptop.
\par
The attacker also needs to start up a DHCP server, DNS proxy, and host the malicious captive portal on his or her laptop. A DHCP server can be set up using the program ISC-DHCP-Server \cite{c15}. The server hands out IP addresses to the victim. The network configuration sent by the DHCP server needs to match the one sent by the legitimate Wi-Fi network exactly. A DNS proxy is set up by using the DnsChef software \cite{c16}. A DNS proxy is needed to resolve URL requests of the victim to the IP address of the attacker's captive portal by applying a DNS spoofing attack. The Apache web server \cite{c17} is used to host the captive portal on the attacker's laptop.
\par
If the victim is already connected to the public Wi-Fi, the attacker can disconnect him or her by initiating a deauthentication attack. This is done by sending continuous deauthentication packets. The program Aireplay-ng \cite{c14} is used to send out these packets. The packets are sent to the victim and the AP of the victim. The packets going to the victim are spoofed as the AP and notify the victim that the AP wants him or her to disconnect from the AP. The packets are also sent to the AP informing it that the victim is disconnecting from the AP. Thus the victim and the AP both disconnect from each other. These packets can target a particular individual if the attacker has the MAC address of the victim, or it can disconnect every individual from the AP.
\par

\begin{figure}[!t]
      \centering
      \includegraphics[scale=0.75]{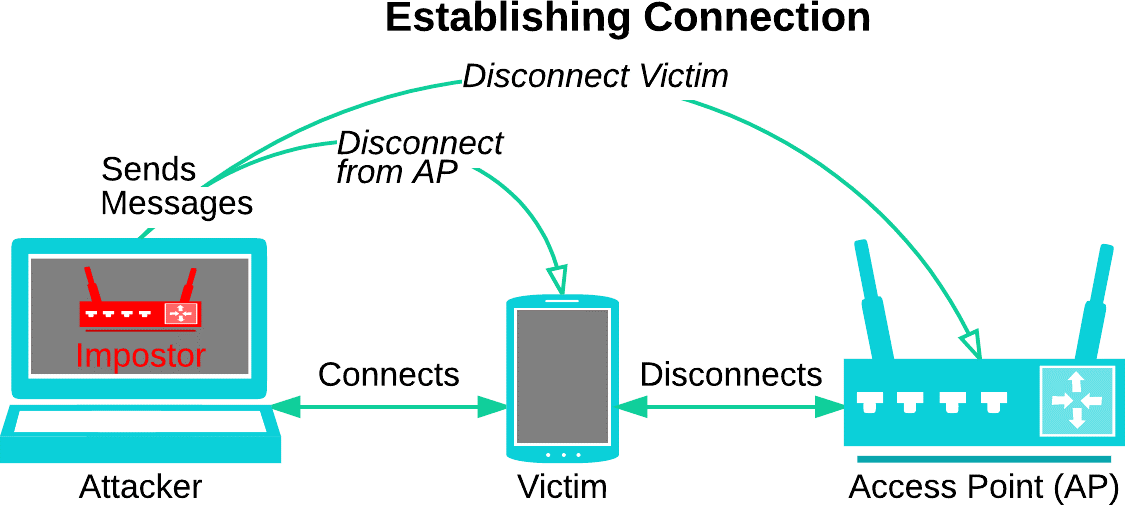}
      \caption{A diagram displaying how the attacker establishes connection with the victim.}
      \label{figure4}
   \end{figure}

Once the victim is disconnected from the AP, he or she would start searching for another AP from the same WiFi network to connect to. As long as the victim has not previously connected to any other nearby network, and as long as the attacker's AP has a stronger signal than the public AP, the victim will connect to the attacker's fake AP. The attacker can power up their wireless card in order to increase the signal strength of their AP to attract the victim. The attacker can also move closer to the victim in order to increase signal strength. Figure~\ref{figure4} illustrates how the attacker establishes a connection with the victim.
\par
\begin{figure}[!t]
      \centering
      \includegraphics[scale=0.75]{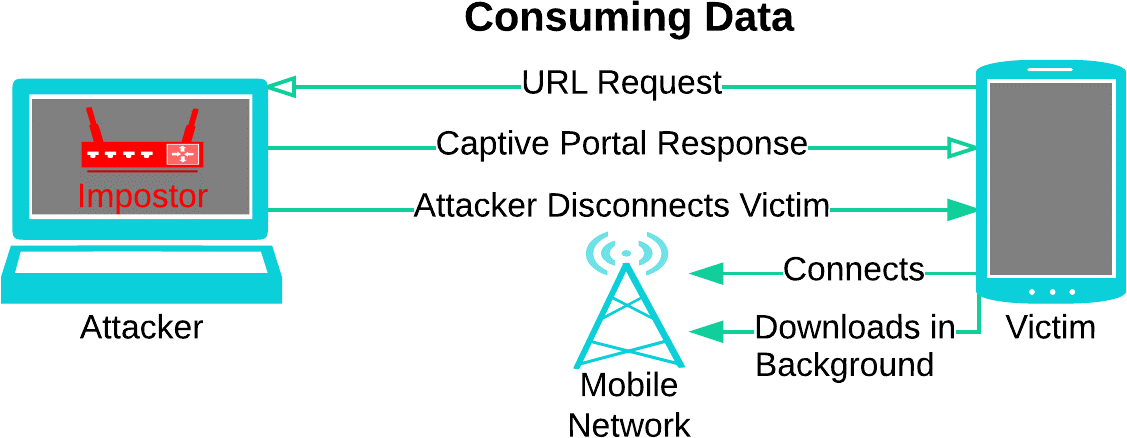}
      \caption{A diagram displaying how the attacker begins consuming data from the victim.}
      \label{figure5}
   \end{figure}
Once the victim connects to the attacker's network, the victim will send an URL request to access a particular website. The attacker DNS proxy would capture the URL request and resolve the request to the IP address of the attacker's captive portal. The victim then requests the IP address which happens to be the malicious captive portal of the attacker. The attacker sends back the captive portal as a response to the victim's request. This is a typical MIMA of the wireless victim.
\par
The victim's browser loads the captive portal and immediately starts trying to download data through the attacker. However, the attacker would not have a connection to the internet, and therefore nothing is yet downloaded by the victim. The victim is expected to click the button on the captive portal that allows a user to connect to the network.
\par
Once the victim has clicked the “the Public Wi-Fi Terms and Conditions” accept button, the attacker disconnects the victim from the Evil Twin Attack access point. This can be done by simply turning off Airebase-ng. The victim now has no nearby networks to connect to, as the attacker begins sending out deauthentication packets as needed to keep the victim from reconnecting to the public AP. With no nearby network to connect to, the victim will switch by default to the mobile data network. The mobile network grants Internet access to the victim and the captive portal. The captive portal now has a connection to download data using the victim's cell phone data plan. At this point, our proposed attack is consuming data from the victim as shown in Figure~\ref{figure5}.

A laptop and online security auditing tools are used in implementing our attack. The attack does not require any modification to a protocol or device firmware. The current implementation of the attack redirects users to a malicious captive portal by poisoning DNS requests.
\par
The attack was tested using two smartphones running Android 6. Both phones used the browser Chrome app to open the captive portal. The rate of the mobile data consumption from the attack varies from one device to another. It also depends on the type of data plan the victim is using. For example, if the victim is enrolled in a high-speed cellular data plan, he or she will consume more data than a slow speed data plan. Other variables to the rate of mobile data consumption include how fast the server pushing data out to the device can do so, and how good of a connection the device has to it's local mobile tower. However, our tests demonstrated that the rate of mobile data consumption was often high enough to cause a severe amount of data consumption.
\par
Our attack can run on a victim's mobile device for an extended time, which will most likely cause a severe amount of data consumption. The attack exploits a vulnerability in the mobile networks' data usage billing system that allows an attacker to cause serious data depletion of customer data quota. Our tests demonstrate that the proposed attack is feasible when a victim connects to a free open Wi-Fi network offered by coffee shops, fast food restaurants, and airports. The attack will keep going as long as the victim does not stop the cellular data connection.
\par
However, the victim may notice that she is not using the Wi-Fi network in two different ways. First, a pop-up message that says “Wi-Fi disconnected” may appear at the bottom of the screen for about a second. Second, an indicator at the top of the screen will show that there is no Wi-Fi connection and that mobile data is being used. The attack attempts to cover indicators through social engineering by showing different messages through the malicious captive portal.
\par
Furthermore, our attack can be only implemented when the victim is connected to an open Wi-Fi network. If the customer is connected to a secure Wi-Fi network that uses WEP/WPA/WPA2, our attack will fail. The attacker will not be able to start an Evil Twin Attack because he or she does not have the wireless network encryption key. Furthermore, the attack may fail when the customer can detect an Evil Twin Attack \cite{c3}\cite{c4}.


\section{CONCLUSIONS} \label{con}
A vulnerability in the mobile networks' data usage billing system was demonstrated by using a mobile data consumption attack. The attack works by delivering a malicious captive portal to the victim, forcing them to connect to their mobile data plan, and causing them to use data via a download initiated by the captive portal. Our attack would work when the victim connects to a free open Wi-Fi network that is available in most coffee shops, fast food restaurants, and airports.
\par
Our attack evaluation was based on attacking the victim for short period of time, using Android mobile OS and Chrome web browser. Further testing is needed to explore the extent of the proposed vulnerability. For example, initiating our attack on different mobile OS and various web browsers. 





\section*{Acknowledgment}
The support for this work was provided by the National Science Foundation REU Site program under Award No. 1560302.
Any opinions, findings, conclusions and recommendations expressed in this material are those of the author(s) and do not necessarily reflect the views of the National Science Foundation.


\end{document}